\documentclass[runningheads]{llncs}
\usepackage[T1]{fontenc}
\usepackage{graphicx}
\usepackage[acronym,toc]{glossaries}
\newacronym{us}{U.S.}{United States}
\newacronym{iot}{IoT}{Internet-of-Things}
\newacronym{rf}{RF}{Radio Frequency}
\newacronym{uart}{UART}{Universal Asynchronous Receiver-Transmitter}
\newacronym{lora}{LoRa}{Long Range Wide Area}
\newacronym{loraw}{LoRaWAN}{Long Range Wide Area Network}
\newacronym{tx}{Tx}{Trasmitter}
\newacronym{rx}{Rx}{Receiver}
\newacronym{rpi3}{RPi4}{Raspberry Pi Model 4b}
\newacronym{sig}{SIG}{Special Interest Group}
\newacronym{ble}{BLE}{Bluetooth Low Energy}
\newacronym{nfc}{NFC}{Near-Field Communication}
\newacronym{rfid}{RFID}{Radio-Frequency Identification}
\newacronym{wifi}{WiFi}{Wireless Fidelity}
\newacronym{i3e}{IEEE}{Institute of Electrical and Electronics Engineers}
\newacronym{lte}{LTE}{Long Term Evolution}
\newacronym{gps}{GPS}{Global Positioning System}
\newacronym{sd}{SD}{Secure Digital}
\newacronym{nmea}{NMEA}{National Marine Electronics Association}
\newacronym{scp}{SCP}{Secure Copy Protocol}
\usepackage{url}
\usepackage{xcolor}
\usepackage{hyperref}
\hypersetup{
 colorlinks,
 linkcolor={red!50!black},
 citecolor={blue!50!black},
 urlcolor={blue!80!black}
}
\usepackage{soul}
\usepackage{lineno}
\modulolinenumbers[1]
\usepackage{float}
\usepackage{subcaption}
\usepackage{algorithm}
\usepackage{algpseudocode}
\begin{document}
\title{SmartAntenna: Enhancing Wireless Range with Autonomous Orientation}
%
%
\author{Michael Swann\inst{1} \and
Pedro Machado\inst{1}\orcidID{0000-0003-1760-3871} \and
Isibor Kennedy Ihianle\inst{1}\orcidID{0000-0001-7445-8573} \and
Salisu Yahaya\inst{1}\orcidID{0000-0002-0394-6112}\and
Farbod Zorriassatine\inst{1}
\and Andreas Oikonomou\inst{1}\orcidID{0000-0002-5069-3971}}
\authorrunning{M. Swann et al.}
%
\institute{Computational Intelligence and Applications Research Group, Clifton Campus, Nottingham Trent University, NG11 8NS UK
\email{n1020368@my.ntu.ac.uk,\{pedro.machado,isibor.ihianle,salisu.yahaya,\\farbod.zorriassatine,andreas.oikonomou\}@ntu.ac.uk}\\
\url{https://www.ntu.ac.uk/research/groups-and-centres/groups/computational-intelligence-applications-research-group}}
\maketitle              
\begin{abstract}
The SmartAntenna proposes a novel approach to extend wireless communication, focusing on autonomous orientation to extend range and optimize performance. Through meticulous evaluation, various aspects of its functionality were assessed, revealing both strengths and areas for improvement. Notably, the antenna tracking mechanism exhibited remarkable efficacy. The SmartAntenna demonstrated robust functionality throughout extensive testing, underscoring its reliability even amidst complex operational scenarios. However, challenges emerged during target tracking, particularly evident in 360-degree sweeps, necessitating further refinement to enhance accuracy. Despite reliance on the HC-12 module, \gls*{lora}, performance limitations surfaced, prompting concerns regarding its suitability for production systems, especially within noisy frequency bands. Nevertheless, the SmartAntenna's adaptability across various wireless technologies holds promise, opening avenues for extended communication ranges and diverse applications. SmartAntenna research contributes valuable insights into optimizing wireless communication systems, paving the way for enhanced performance and expanded capabilities in diverse operational environments.

\end{abstract}
\section{Introduction}\label{ch:intro}
In 1896, Guglielmo Marconi, inspired by Heinrich Hertz's pioneering work on radio waves, secured a patent for the world's inaugural wireless telegraph \cite{granatstein2007}. His groundbreaking experiment in 1901, spanning from Cornwall, UK to Newfoundland, Canada, marked a historic achievement in communication \cite{huang2008}. The success of the transmission sent ripples of concern through the established cable telegraph companies of the era, which had heavily invested in underwater cabling connecting Europe and the \gls*{us} \cite{granatstein2007}. Over a century later, the electromagnetic waves central to Marconi's innovation continue to underpin some of today's most sophisticated wireless technologies. As recognised by the cable telegraph companies of Marconi's time, these wireless systems offer numerous advantages over their wired counterparts, including mobility, cost-effective coverage of expansive areas, and, in contemporary contexts, data rates that rival those of wired connections \cite{celebi2020}. Wireless communication still presents various challenges such as signal weakening over distances, interference from obstacles, and regulatory constraints on radio frequencies \cite{beard2015}. Despite these hurdles, multiple wireless solutions exist, each prioritising different features like data rates, energy efficiency, or range \cite{celebi2020}. For instance, while \gls*{wifi} offers high data rates but consumes significant power, Bluetooth excels in security and energy efficiency but has limited range, and the HC-12\footnote{Available online, \protect\url{https://statics3.seeedstudio.com/assets/file/bazaar/product/HC-12_english_datasheets.pdf}, last accessed: 15/05/2024}, a \gls*{lora} device provides extensive range at the expense of transmission rates. \gls*{lora} transceivers have gained traction in wireless systems, particularly in applications that require lower transmission rates, such has \gls*{iot}. However, discrepancies arise between claimed and actual performance, with reported ranges varying significantly \cite{marpaung2020,winasis2022}. Moreover, factors like latency are crucial for applications like robotic control \cite{celebi2020}. The SmartAntenna project pioneers an innovative autonomous antenna orientation system, prioritising critical facets of wireless communication: latency, throughput, and range. The innovative technology holds promise for deployment across diverse and demanding environments, including underwater, mining, and offshore operations. Moreover, it could be instrumental for emergency response services, facilitating the extension or restoration of compromised wireless communication infrastructures in the aftermath of natural disasters, conflicts, or other natural crises such as wildfires, floods and storms.
The structure of the article is as follows: the relevant research is covered in Section \ref{ch:lr}, the design and implementation of the SmartAntenna are detailed in Section \ref{ch:methodology}, the analysis of results is conducted in Section \ref{ch:results}, and the final conclusions and prospects for future work are addressed in Section \ref{ch:conclusion}.

\section{Background research}\label{ch:lr}
Wireless communication technologies play a pivotal role in \gls*{iot} systems, facilitating data transmission wirelessly between devices across varying distances. The selection of a wireless technology often depends on the specific application domain of the \gls*{iot} system, ranging from short-range connections to long-distance transmissions. The market's heterogeneity and the variety of available technologies necessitate categorisation based on factors such as electromagnetic wave length and operating frequencies. These technologies encompass Radio Frequency Transmission, Infrared Transmission, Microwave Transmission, and Lightwave Transmission \cite{sikimic2020}. Bluetooth, developed by the Bluetooth \gls*{sig}, primarily facilitates short-range communication between devices, particularly prevalent in mobile and computer peripheral connections. \gls*{ble} emerged as a variant, offering reduced power consumption and setup time, catering to devices with low bandwidth requirements \cite{sikimic2020}. \gls*{nfc}, engineered by Sony and Philips, enables two-way data transfer between devices placed in close proximity, typically within a 10 cm range. Widely utilised in mobile devices for contactless payment and data access, \gls*{nfc} operates on magnetic coupling principles and \gls*{rfid} technology \cite{curran2012}. \gls*{wifi}, standardised by the \gls*{i3e}) as 802.11, represents a ubiquitous communication standard renowned for its broad coverage, high data transfer rates, and extensive market penetration. Utilising frequencies in the 2.4 to 5 GHz range, \gls*{wifi} requires an access point and wireless network adapter for network establishment \cite{song2014}. Cellular networks, encompassing generations from 1G to 5G \gls*{lte}, form large-scale communication infrastructures divided into cells, each serviced by a transceiver. Characterised by low energy consumption and expansive coverage, cellular networks cater to high-bandwidth requirements and are prevalent in mobile phones and \gls*{iot} devices. The latest iterations, 5G and \gls*{lte}, offer enhanced speed, capacity, and device connectivity \cite{manam2019}.
\gls*{lora} and \gls*{loraw} represent distinct components within the realm of wireless communication \cite{shah2016}. \gls*{lora} serves as the physical layer technology developed by Semtech, enabling long-distance wireless communication. It operates by modulating data onto radio waves, offering robust performance even in challenging environments. On the other hand, \gls*{loraw} encompasses the higher layers of the communication stack, serving as a protocol for managing communication between \gls*{lora}-equipped devices and network gateways. \gls*{loraw} defines the network architecture, addressing, and data rates, facilitating efficient and scalable communication over \gls*{lora}'s physical layer. While \gls*{lora} provides the underlying technology for long-range communication, \gls*{loraw} adds the necessary network infrastructure and protocols for creating wide-area \gls*{iot} networks, enabling devices to securely transmit data over long distances to centralised network servers \cite{shah2016}. The \gls*{loraw} protocol, compatible with European and North American regulatory standards, operates on specific frequency bands facilitating data transmission \cite{shah2016}. Many commercial and industrial solutions leverage \gls*{loraw}, communicating with backend systems through network and application servers, formerly named the \gls*{lora} Network and \gls*{lora} Application Server \cite{shah2016}. In contrast, NB-\gls*{iot}, belonging to the \gls*{lte}-\gls*{iot} network category, operates on 4G mobile networks, offering extensive coverage and reduced power consumption. ZigBee, standardised by \gls*{i3e} in 2003, remains distinguished for its longevity, often exceeding 10 years, and is commonly used in sensor networks owing to its reliability and cost-effectiveness, albeit with a limited transmission distance of up to 100 meters \cite{sikimic2020}. For the SmartAntenna project, it was decided to focus on \gls*{lora}.
The HC-12 is a wireless \gls*{rf} communication module is a \gls*{lora} transceiver that operates in the 433.4 – 473.0 MHz range. It uses a \gls*{uart} connection to communicate with a host device and can be installed in place of a wired \gls*{uart} connection without any adjustments to the system software. The devices are relatively cheap. Devices used in the SmartAntenna were roughly £3.20 but can be lower if bought in bulk. Another useful feature of the device is that it can be configured in various modes that effect range, power consumption, and data rate. 
Marpaung et al. \cite{marpaung2020} developed an early warning system for peat land fires that relied on the HC-12 to transmit temperature readings from remote devices to a base device. They stated in their description of the HC-12 that it has a maximum range of 1800 meters, while in their testing achieved a range of 870 meters in the long-range setting (FU4) and 460 meters in the default mode (FU3). A prototype wearable device that utilises the HC-12 was developed by Hassaballah et al. \cite{hassaballah2020} for the medical industry, to enable doctors to wirelessly monitor patients. The accompanying report stating that the devices small size and ease of use have made it popular and the frequency band it uses is safe for use in the medical industry. Bhattacharyya et al. \cite{bhattacharyya2022} found the HC-12 module, used in conjunction with an Arduino Mega 2560\footnote{Available online, \protect\url{https://docs.arduino.cc/hardware/mega-2560/}, last accessed: 15/05/2024}, was very suitable and cost-effective for the task of connecting Gas Electron Multiplier detectors. Bhattacharyya et al. \cite{bhattacharyya2022} mentioned that the system only required a data rate of 40 characters roughly every 30 seconds. Winasis et al. \cite{winasis2022} developed a prototype device for assisting the deaf during prayer time, sending a signal to a vibrating wrist device when they were expected to follow certain movements. Although they claim the HC-12 improved range performance over a previous version, the maximum range achieved for their solution was only 30 metres and tests beyond that were unsuccessful. Nevertheless, all of the revised works share in common the ambiguity surrounding the range performance. For instance, the 30 metres stated by Winasis et al. \cite{winasis2022} contradict the 870~1000 meters claimed by Marpaung et al. \cite{marpaung2020}. What is also not clear is how bandwidth and latency performance effected these projects. None of these applications seem to be particularly time-critical nor transmitting substantial amounts of data. It is, however, important for the SmartAntenna project and other future projects that wish to utilise the HC-12, to know accurately what the performance is of the device. As a clear answer has not been obtained from the literature review, it would be necessary to perform tests on the device.

\section{Methodology}\label{ch:methodology}
To facilitate testing, two devices were developed that utilised the HC-12 module. A base device consisting of a \gls*{rpi3}\footnote{Available online, \protect\url{https://www.raspberrypi.com/products/raspberry-pi-4-model-b/}, last accessed: 15/05/2024} which could be powered by mains electricity or a battery pack and a portable device using an Arduino Nano that was battery powered. Tests for throughput, latency, and range were conducted. As the HC-12 can be configured in various modes, a number of these would be tested to see how they affected performance. For the latency and throughput tests, a wired connection was also tested for comparison. Furthermore, the HC-12 connects to the \gls*{tx} and \gls*{rx} pins of the \gls*{uart} port.
\begin{figure}[h]
    \vspace*{-5mm}

	\centering
	\includegraphics[scale=0.25]{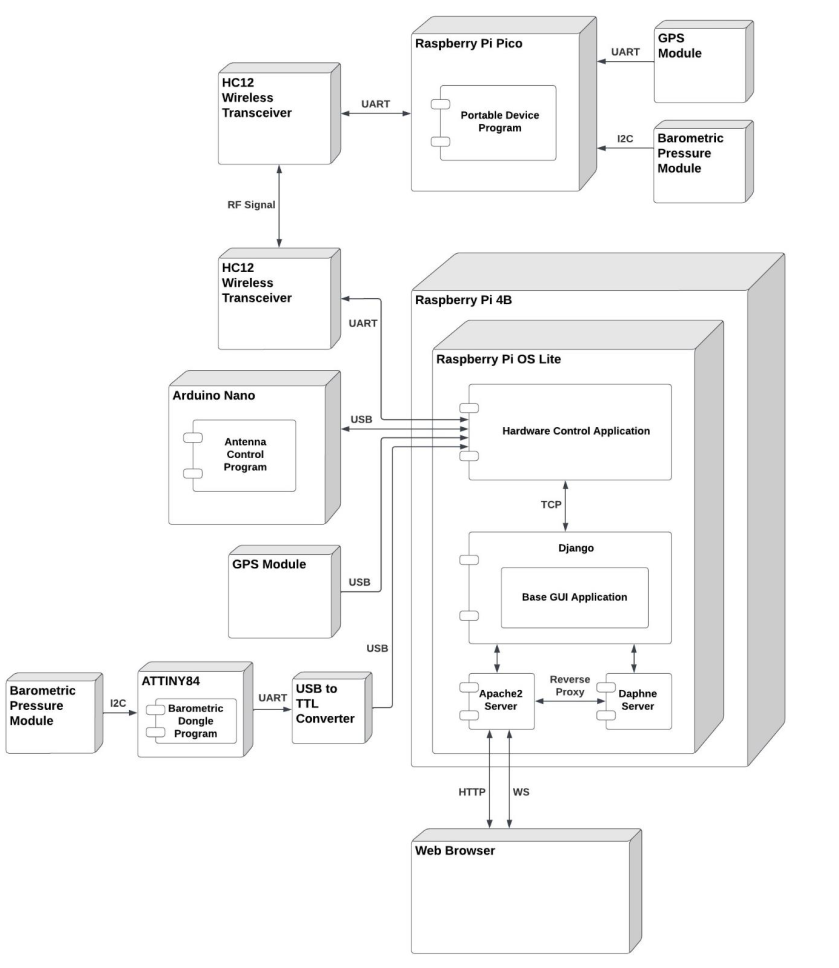}
	\caption{Deployment diagram illustrating the various hardware and software
components of the proposed system.}
 \label{fig:diagram}
    \vspace*{-5mm}
\end{figure}

Fig. \ref{fig:diagram} showcases a system designed for real-time communication and data collection. At its core lies the Base Unit, a powerful \gls*{rpi3} that acts as the central hub for communication and data processing. ZMQ, a high-performance messaging library, facilitates seamless data exchange between the Base Unit and other system components. ZMQ ensures smooth communication throughout the system. Web Sockets enable real-time, two-way communication between the Base Unit and a web server. Web sockets enable constant data exchange and updates, crucial for remote monitoring and control via a web browser. The Web UI, a user interface developed using Django, provides users with a web-based interface for interacting with the system. Users can view real-time sensor data (potentially from the portable device), track the current antenna position, and send control commands to both the antenna and the portable device. The system boasts an Offline Mode, ensuring functionality even without an internet connection. Enabling the reliable operation in diverse environments where internet access might be limited. The Barometric Dongle, an external device, gathers atmospheric pressure data. The data can be used for various purposes, such as providing additional environmental context for sensor readings or enhancing the accuracy of other sensors. Another external device, the GPS Dongle, supplies real-time location information. \gls*{gps} is essential for accurate tracking and positioning of the antenna relative to the portable device. The system operates through a two-way data flow. Sensor data from the portable device likely travels through web sockets to the Base Unit. The Base Unit then processes the data and presents it for real-time monitoring through the web UI. Control commands issued from the web UI are likely transmitted via web sockets to the Base Unit, which then relays these commands to the antenna and/or portable device. 

To measure the amount of throughput the device is capable of, the Arduino device would print 10 Kb of data to the serial output. A series of characters at the start and end of the data would signal the receiver to record the time. The same data would be used every time and had been saved to a text file on the \gls*{rpi3}, so it could be compared with the data received. Algorithm \ref{alg:start_cond} shows the pseudocode for the receiving device.
\vspace*{-5mm}
\begin{algorithm}
\caption{Pseudocode for detecting start condition and reading bytes}\label{alg:start_cond}
\begin{algorithmic}[1]
\While{start condition not detected}
    \State Listen for start condition
    \State SET counter TO 0
    \State CREATE buffer SIZE 10001
    \State SET start TO system clock now
    \While{TRUE}
        \State SET $c$ to read byte from serial
        \State ADD $c$ to buffer
        \State INCREMENT counter
        \If{end condition detected OR counter $\geq$ 10000}
            \State \textbf{BREAK}
        \EndIf
        \State SET stop TO system clock now
        \If{buffer EQUALS testData}
            \State PRINT difference between stop and start
        \Else
            \State PRINT test failure message
        \EndIf
    \EndWhile
\EndWhile
\end{algorithmic}
\end{algorithm}
\vspace*{-5mm}
Another important performance metric is latency. To measure latency, a device would send a single character (1 byte) to another device, which would immediately return the character to the sender. The sending device would time how long it took from sending the character to receiving it. The pseudocode for the sending device can be seen in Algorithm \ref{alg:start_cond1}. Again, the tests were first conducted with a wired connection to help isolating the additional time penalties caused by the processor on the HC-12 module.
\begin{algorithm}
\caption{Pseudocode for sending data over serial in a loop}\label{alg:start_cond1}
\begin{algorithmic}[1]
\For{$i$ \textbf{from} 0 \textbf{to} 9}
    \State Flush rx and tx buffers
    \State SET start TO system clock now
    \State Send one byte of data over serial
    \While{no serial available}
        \State Do nothing
    \EndWhile
    \State SET received TO serial read byte
    \State SET stop TO system clock now
    \If{received EQUALS sent byte}
        \State Store difference between stop and start
    \EndIf
\EndFor
\end{algorithmic}
\end{algorithm}
To evaluate range performance a test was devised that required the portable device to transmit its current \gls*{gps} location to the base device. The coordinates of the furthest transmission received could then be used to measure the distance. The portable device consisted of an \gls*{sd} card reader, \gls*{gps} module, and HC-12 module connected to an Arduino Nano. The \gls*{gps} module would receive new coordinates every second. This would be increased to a transmission at 10 second intervals (later 5 seconds), due to the time needed to process the \gls*{gps} data. At the end of each interval, if the \gls*{nmea} sentence was valid then it was written to the \gls*{sd} card. If the write command successful, then it is transmitted via the HC-12 module.

The reason for validating the coordinates before transmitting them was because the \gls*{nmea} sentence checksum would be used later to confirm successful transmission. If the portable device were to transmit invalid messages, which would be picked up as an error in the transmission from the HC-12, when it is in fact an error from the \gls*{gps} module. It was also important to log transmissions into the \gls*{sd} card on the sending device to record exactly what was sent. If an absence of transmissions was tested at the receiver only, then the performance of the HC-12 could again be affected by other components. For example, the Arduino freezing, or power loss, would look to the receiver as if a transmission had been lost. A python program was developed to process the \gls*{nmea} sentences from both the \gls*{sd} card and retrieve the file from the base device using \gls*{scp}. Coordinates that were present on both devices were plotted on a map as successful transmissions, and coordinates only present on the portable device plotted as lost transmissions.

The antenna development involved creating a controllable unit based on a gimbal design, actuated by a positional servo and a motor. The Arduino Nano was chosen as the microcontroller due to its small form factor and ease of communication via USB. PlatformIO was used for programming, leveraging its cross-platform capabilities and integration with VSCode for enhanced development features. The servo's speed was controlled via software to avoid mechanical wobble, enhancing stability during operation.

\section{Results Analysis}\label{ch:results}
\begin{figure}[h]
    \vspace*{-5mm}
    \centering
    \begin{tabular}{cc}
        \begin{minipage}{0.45\textwidth}
            \centering
            \includegraphics[scale=0.12]{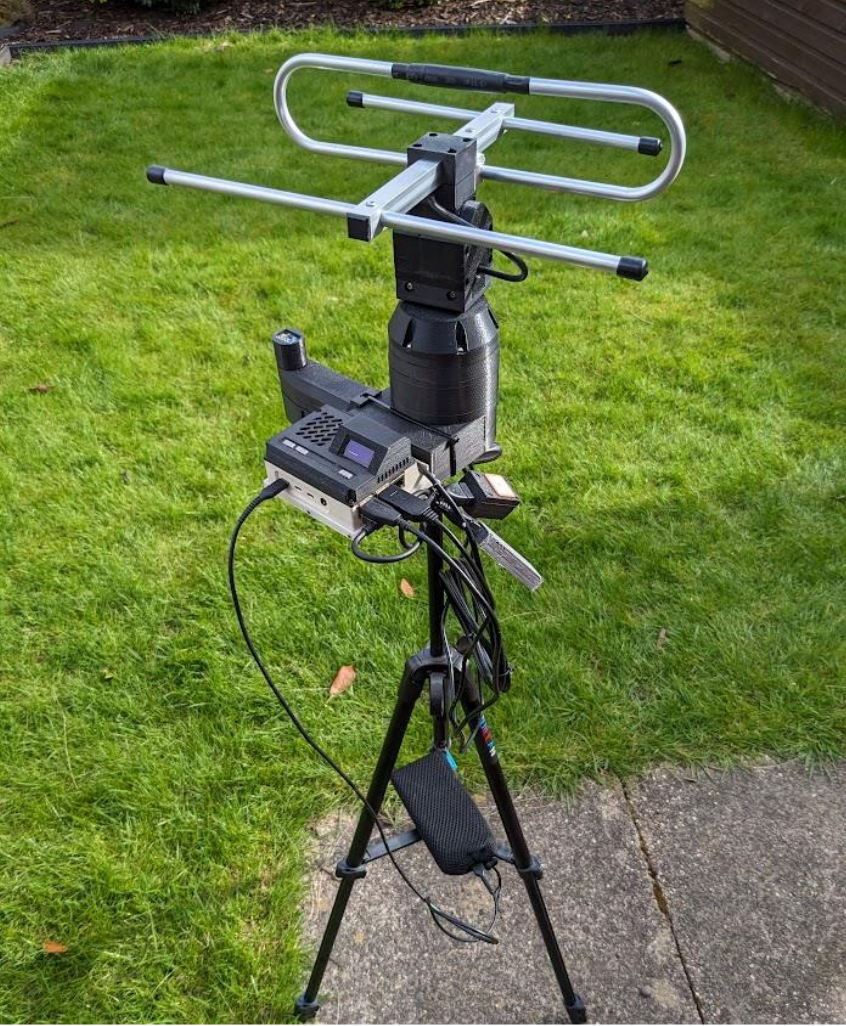} 
            \caption{Portable device powered by Raspberry Pi Pico}
 \label{fig:base}
        \end{minipage}
        &
        \begin{minipage}{0.45\textwidth}
            \centering
            \includegraphics[scale=0.25]{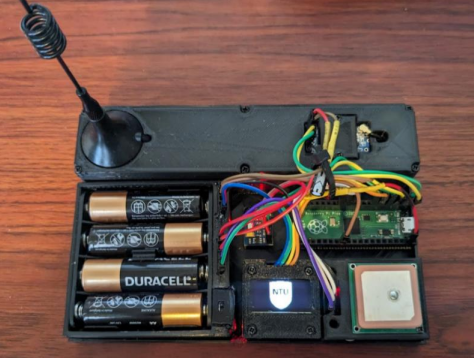}
            \caption{Antenna system comprising of tracking antenna, base control unit,
and positional USB devices.}\label{fig:portable}
        \end{minipage} \\
    \end{tabular}
    \vspace*{-5mm}
\end{figure}

The implemented SmartAntenna system was composed of the base unit (see Fig. \ref{fig:base}) and a portable unit (see Fig. \ref{fig:portable}).

The results in Table \ref{tab:throughput} show the wired results are only slightly slower than the calculated minimum for each baud rate, which proves the test program was behaving correctly. The test was then performed on the wireless configurations chosen using the HC-12. Some configurations that were tested use the same output baud rate but with differing modes which can affect performance.

\begin{table}
    \vspace*{-10mm}
\centering
\caption{Results from throughput test measuring the time taken to transfer 10Kb of data.}
\label{tab:throughput}
\begin{tabular}{|c|c|c|c|c|c|c|}
\hline
ID & Baud rate & Mode & Time taken [s] & Wired equivalent time [s] & Min. Calculated \\
\hline
1 & 1200 & FU4 & 194.436825 & 83.649479 & 83.33333333 \\
2 & 2400 & FU3 & 41.884853 & 41.877711 & 41.66666667 \\
3 & 4800 & FU2 & 50.904906 & 21.011591 & 20.83333333 \\
4 & 9600 & FU1 & 10.547779 & 10.559831 & 10.41666667 \\
5 & 9600 & FU3 & 10.560724 & 10.559831 & 10.41666667 \\
6 & 38400 & FU3 & 2.701640 & 2.700745 & 2.604166667 \\
7 & 115200 & FU1 & 0.924378 & 0.931097 & 0.868055556 \\
8 & 115200 & FU3 & 0.938599 & 0.931097 & 0.868055556 \\
\hline
\end{tabular}
\vspace*{-5mm}
\end{table}

Most have a very similar result to the wired equivalent, meaning the HC-12 is mostly not impeding the transfer of data. The bottleneck for data transfer in most cases seems to be the UART connection with the device. Tests 1 and 3 are the only results with significant time differences to the wired test. The modes used in these tests prioritise low power and long range respectively. They differ from the other modes in that the data transfer over the air is slower than the \gls*{uart} connection speed. If data is continuously sent to the device in these modes, the input buffers will overflow, and data will be lost. The HC-12 user manual suggests sending packets no bigger than 20 bytes for FU2 and 60 bytes for FU4 mode, with >2 seconds between each packet. In some of the tests the wired connection was very slightly slower, which seems like it should be impossible. It is possible that the length of the signal wires in the wired connection may have influenced the result due to their higher capacitance. The difference is less than 1\% in any case, so it is not of significant concern.

The latency test results in Table \ref{tab:latency} indicate that there are additional overheads created by the device.
\begin{table}
\centering
\vspace*{-10mm}
\caption{Averaged results of latency test on different modes and baud rates.}
\label{tab:latency}
\begin{tabular}{|c|c|c|c|c|}
\hline
ID & Baud rate & Mode & Avg response (ms) & Wired equivalent (ms) \\
\hline
1 & 1200 & FU4 & 2592.411621 & 14.996183 \\
2 & 2400 & FU3 & 189.5862802 & 7.5204258 \\
3 & 4800 & FU2 & 524.0615822 & 3.8166278 \\
4 & 9600 & FU1 & 32.6835997 & 1.9512919 \\
5 & 9600 & FU3 & 86.7270886 & 1.9512919 \\
6 & 38400 & FU3 & 35.6899399 & 0.5407718 \\
7 & 115200 & FU1 & 30.6233627 & 0.244038 \\
8 & 115200 & FU3 & 15.8306192 & 0.244038 \\
\hline
\end{tabular}
\vspace*{-5mm}
\end{table}
Again, tests using the FU2 and FU4 modes are showing the most significant difference from the wired speeds. Tests 4 and 7 using mode FU1 have a very similar time despite very different baud rates. In the user manual it states in mode FU1, the air baud rate is a uniform 250,000 bps regardless of the \gls*{uart} baud rate. However, in mode FU3 the air baud rate changes with the \gls*{uart} baud rate, explaining why these speeds vary in comparison.

The farthest distance from a successful transmission of each mode tested is shown in Table \ref{tab:hc12_range_test}.
\begin{table}
\centering
\vspace*{-10mm}
\caption{HC-12 Range test results of various modes.}
\label{tab:hc12_range_test}
\begin{tabular}{|c|c|c|c|c|}
\hline
Baudrate & Mode & Furthest distance (metres) & Test ID \\
\hline
1200 & FU4 & 60.84 & 184 \\
4800 & FU2 & 9.15 & 187 \\
9600 & FU1 & 12.14 & 185 \\
9600 & FU3 & 37.55 & 255 \\
38400 & FU3 & 17.6 & 180 \\
115200 & FU3 & 13.85 & 182 \\
\hline
\end{tabular}
\vspace*{-5mm}
\end{table}

The range tests on test site were conducted up to 100 metres and it was expected that most configurations would manage the distance. The results of the tests revealed that none of the modes achieved the 100 metres and the maximum experimental range was 60 metres in the long-range mode (FU4). Unfortunately, not all configurations could be tested due to issues in the field that could not be resolved. Given that the long-range mode and the default mode were successfully tested, it was not necessary to conduct these tests again.

The most surprising result of these tests is the range in all modes. A maximum of 60.84
metres is significantly less than what has been claimed. The results also show that modes
FU2 and FU4 have too high a latency to be really considered for the application. Mode FU1
was even poorer range and as power consumption is not a priority for the project it is not
worth considering. Mode FU3 at 9600 baud rate has the most balanced performance and
most likely why it is the default mode. Lowering the baud rate in the mode also lowers
the air baud rate, leading to an increased range. Lower baud rates in FU3 mode could be
considered to extend the range, if it is found that data transfer speeds are not as important.

\begin{figure}[h]
    \vspace*{-5mm}

    \centering
    \begin{tabular}{cc}
        \begin{minipage}{0.45\textwidth}
            \centering
            \includegraphics[scale=0.1]{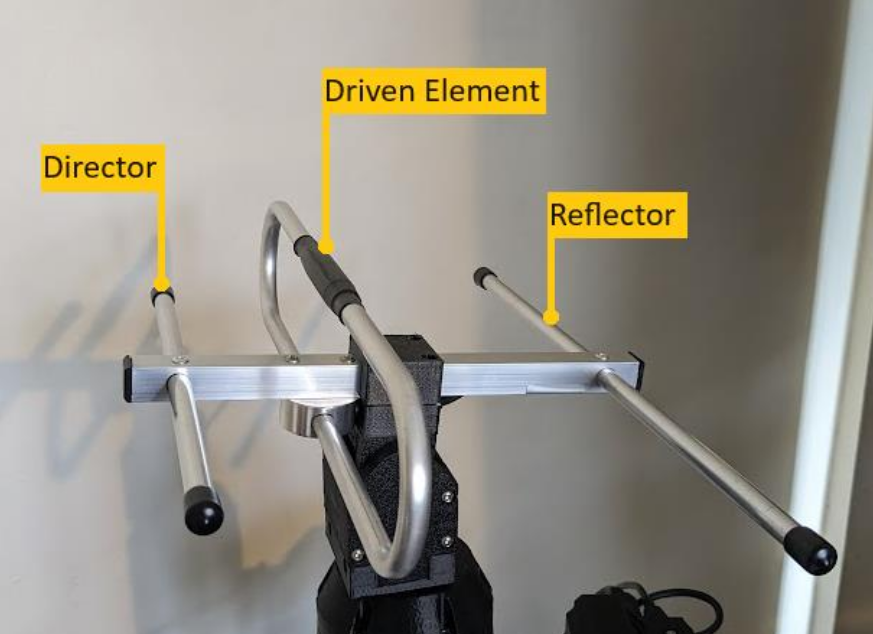} 
            \caption{433MHz Yagi Uda antenna}
 \label{fig:antenna1}
        \end{minipage}
        &
        \begin{minipage}{0.45\textwidth}
            \centering
            \includegraphics[scale=0.11]{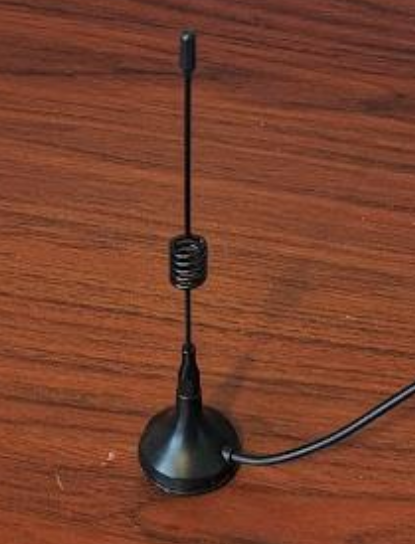}
            \caption{Dolla Tek Unity Gain omnidirectional antenna.}\label{fig:antenna2}
        \end{minipage} \\
    \end{tabular}
    \vspace*{-10mm}
\end{figure}

Two more antennas were purchased, which could potentially improve range performance. The Unity Gain omnidirectional antenna and a Yagi Uda directional antenna (see Fig. \ref{fig:antenna1}). The Unity Gain antenna works the same way as the previous antenna, just with a much longer antenna length (see Fig. \ref{fig:antenna2}). A Yagi-Uda antenna has the potential to drastically increase the range of a radio signal by utilising additional components. As well as the driven element or dipole, there also exists a reflector element that sits behind the dipole and a director element in front of it. The director and reflector are not electrically connected to the dipole and are known as parasitic elements \cite{huang2008}. The way these parasitic elements are excited by incoming radio waves causes them to emit electromagnetic radiation, and when arranged correctly, the radiation from these parasitic elements can strengthen the signal as it reaches the dipole \cite{hartnagel2023}. 
Table \ref{tab:range_test} shows the results from each antenna. The Yagi Uda antenna was also tested while pointing perpendicular to the receiver, to show the directional effect. It should also be noted that for the Yagi Uda tests, the Unity Gain antenna was used as the antenna for the receiver.
\begin{table}
\vspace*{-10mm}
\caption{Range tests conducted on two antennas designed for increased range performance.}
\label{tab:range_test}
\begin{tabular}{|c|c|c|c|c|}
\hline
Baud rate & Mode & Furthest distance (metres) & Test ID & Antenna \\
\hline
9600 & FU3 & 131.1 & 344 & Unity Gain omni-directional \\
9600 & FU3 & 137.95 & 345 & Yagi Uda \\
9600 & FU3 & 46.99 & 346 & Yagi Uda (aimed perpendicular) \\
\hline
\end{tabular}
\vspace*{-7mm}
\end{table}
The results show that the range is indeed increased by a substantial amount by utilising these antennas. The furthest distance for the Unity Gain antenna is slightly misleading, which can be seen by studying the \gls*{gps} results shown in Appendix A. Although the unity gain antenna reached 131.1 metres, it only had consistently successful transmissions up to about 90 metres. What should also be noted, the Yagi Uda antenna reached 137.95 metres, but that is the maximum tested distance and it may have reached further as no signals were dropped in the entire test. To demonstrate the directional properties of the Yagi Uda antenna another test with the antenna aiming perpendicular to the receiver which yielded much worse results.
The range improvements achieved by the Yagi-Uda antenna were deemed good enough for the project to continue with the HC-12 module as the wireless technology. There is an obvious drawback with using a directional antenna to communicate with a moving receiver, however. Therefore, the moving target will move out of the optimal signal strength. The solution involves complex algorithms to calculate the direction-of-arrival of a signal and then electronically adapt the radiation pattern of the antenna array to maximise the signal (Bazan, Kazi and Jaseemuddin, 2021). But as well as being very difficult to implement, the HC-12 does not provide any measurement of signal strength that could be used in the algorithms. Another solution is to physically move a directional antenna at the target, known as a tracking antenna. There is another similar design by Airborne innovations\footnote{Available online, \protect\url{https://www.airborneinnovations.com/ai/datalinks/tracking-antennas/}, last accessed: 18/05/2024}, who do not explicitly say how the antenna tracks but state that it has integrated \gls*{gps} and compass. Their design also includes an additional omni-directional antenna for close range operations. A tracking antenna seems like a more viable option and as \gls*{gps} had already been implemented successfully for range testing.

\section{Conclusions and Future Work}\label{ch:conclusion}
The antenna tracking mechanism demonstrated unexpected efficacy, notably outperforming other motors despite being 3D printed. Despite comprising numerous components, the system exhibited robust functionality without notable issues during extensive testing sessions. However, challenges arose during target tracking, particularly with 360-degree sweeps, resulting in some inaccuracies. Yet, the single-director antenna utilised displayed adequate accuracy for its purpose. Latency, averaging 253.05ms and peaking at 505ms, while acceptable for many applications, proved slower for robotic control, warranting optimisation for more direct command signal transmission. However, the SmartAntenna's adaptability for various wireless technologies offers promising prospects, particularly if paired with a more reliable wireless module. Additionally, hardware selections such as the Raspberry Pi Pico and \gls*{gps} receivers showcased commendable performance and affordability, underscoring their suitability for projects requiring positioning accuracy and microcontroller capabilities. Moreover, deploying a web application on the \gls*{rpi3} proved an effective solution for scalable interface access, albeit requiring consideration for remote deployment in areas lacking Wi-Fi access, suggesting potential integration of touch screen functionality for improved usability in future iterations.

Future work for the SmartAntenna entails addressing performance limitations associated with the HC-12 module, which raise concerns about its suitability for production systems due to issues and potential defects, especially within the noisy 433MHz band. Suggestions for enhancement include integrating an omnidirectional antenna alongside the directional one to mitigate positional errors at close ranges, as proposed by Airborne Innovations. Moreover, the SmartAntenna's potential application in extending aerial, terrestrial, or underwater communications range at a reasonable cost presents an intriguing use case. 

\begin{credits}
\subsubsection{\discintname}
The authors have no competing interests to declare that are
relevant to the content of this article.
\end{credits}
%
%
%
%
\addcontentsline{toc}{section}{References}
\bibliographystyle{splncs04}
\bibliography{references}

\begin{thebibliography}{10}
\providecommand{\url}[1]{\texttt{#1}}
\providecommand{\urlprefix}{URL }
\providecommand{\doi}[1]{https://doi.org/#1}

\bibitem{beard2015}
Beard, C., Stallings, W.: Wireless communication networks and systems. Pearson (2015)

\bibitem{bhattacharyya2022}
Bhattacharyya, R., Sahu, S., Sahu, P.: Remote data acquisition with wireless communication for a quad-gem detector. Journal of Instrumentation  \textbf{17}(02),  T02001 (2022)

\bibitem{celebi2020}
Celebi, H.B., Pitarokoilis, A., Skoglund, M.: Wireless communication for the industrial iot. Industrial IoT: Challenges, Design Principles, Applications, and Security pp. 57--94 (2020)

\bibitem{curran2012}
Curran, K., Millar, A., Mc~Garvey, C.: Near field communication. International Journal of Electrical and Computer Engineering  \textbf{2}(3), ~371 (2012)

\bibitem{granatstein2007}
Granatstein, V.L.: Physical principles of wireless communications. Auerbach publications (2007)

\bibitem{hartnagel2023}
Hartnagel, H.L., Quay, R., Rohde, U.L., Rudolph, M.: Fundamentals of RF and Microwave Techniques and Technologies. Springer (2023)

\bibitem{hassaballah2020}
Hassaballah, H.J., Fayadh, R.A.: Implementation of wireless sensor network for medical applications. In: IOP conference series: materials science and engineering. vol.~745, p. 012089. IOP Publishing (2020)

\bibitem{huang2008}
Huang, J., Encinar, J.A.: Reflectarray antennas, a john wiley \& sons. Inc., Publication  (2008)

\bibitem{manam2019}
Manam, S., Yashwanth, P., Telluri, P.: Comparative analysis of digital wireless mobile technology: a survey. International Journal of Innovative Technology and Exploring Engineering  \textbf{8},  268--273 (2019)

\bibitem{marpaung2020}
Marpaung, N., Amri, R., Ervianto, E.: Analysis of wireless fire detector application to detect peat land fire based on temperature characteristic. In: IOP Conference Series: Materials Science and Engineering. vol.~846, p. 012051. IOP Publishing (2020)

\bibitem{shah2016}
Shah, S.H., Yaqoob, I.: A survey: Internet of things (iot) technologies, applications and challenges. 2016 IEEE Smart Energy Grid Engineering (SEGE) pp. 381--385 (2016)

\bibitem{sikimic2020}
Sikimić, M., Amović, M., Vujović, V., Suknović, B., Manjak, D.: An overview of wireless technologies for iot network. In: 2020 19th International Symposium INFOTEH-JAHORINA (INFOTEH). pp.~1--6 (2020). \doi{10.1109/INFOTEH48170.2020.9066337}

\bibitem{song2014}
Song, S., Issac, B.: Analysis of wifi and wimax and wireless network coexistence. arXiv preprint arXiv:1412.0721  (2014)

\bibitem{winasis2022}
Winasis, M.M.: PRAYER GUIDE TOOL FOR DEAF USING GYROSCOPE SENSOR AND HC-12. Ph.D. thesis, Universitas Muhammadiyah Yogyakarta (2022)

\end{thebibliography}

\end{document}